# Metal to insulator quantum-phase transition in few-layered ReS$_2$

Nihar R. Pradhan[†], Amber McCreary[∥‡§], Daniel Rhodes[†,¤], Zhengguang Lu[†,¤], Simin Feng[∥], Efstratios Manousakis[†,¤], Dmitry Smirnov[†], Raju Namburu[‡], Madan Dubey[§], Angela R. Hight Walker[‡], Humberto Terrones[ι], Mauricio Terrones[∥,ϕ], Vladimir Dobrosavljevic[†,¤] and Luis Balicas[†,*]

[†]National High Magnetic Field Lab, Florida State University, 1800 E. Paul Dirac Dr. Tallahassee, FL 32310, USA.
[∥]Department of Physics and Center for 2-Dimensional and Layered Materials, Pennsylvania State University, Pennsylvania, PA 16802, USA.
[‡]Computational & Information Sciences Directorate, U.S. Army Research Laboratory, Adelphi, MD 20783, USA.
[§]Sensor & Electron Devices Directorate, U.S. Army Research Laboratory, Adelphi, MD 20783, USA.
[¤]Department of Physics, Florida State University, Tallahassee, Florida 32306, USA.
[‡]Optical Technology Division, Physics Laboratory, National Institute of Standards and Technology, Gaithersburg, MD 20899, USA.
[ι]Department of Physics, Applied Physics and Astronomy, Rensselaer Polytechnic Institute, 110 Eighth Street, Troy, New York 12180, USA.
[ϕ]Department of Chemistry and Department of Materials Science and Engineering, The Pennsylvania State University, University Park, PA 16802, USA & Carbon Institute of Science and Technology, Shinshu University, Wakasato 4-17-1, Nagano-city 380-8553, Japan.

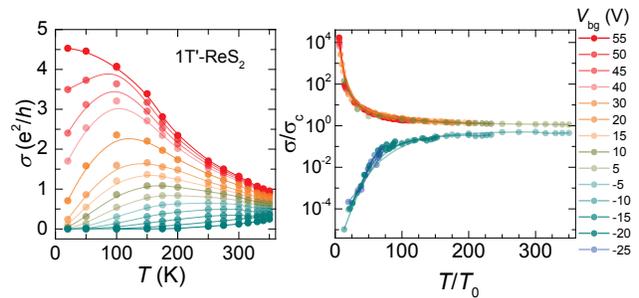

In ReS$_2$ a layer-independent direct band-gap of 1.5 eV implies a potential for its use in optoelectronic applications. ReS$_2$ crystallizes in the 1T'-structure which leads to anisotropic physical properties and whose concomitant electronic structure might host a non-trivial topology. Here, we report an overall evaluation of the anisotropic Raman response and the transport properties of few-layered ReS$_2$ field-effect transistors. We find that ReS$_2$ exfoliated on SiO$_2$ behaves as an $n$-type semiconductor with an intrinsic carrier mobility surpassing $\mu_i \sim 30$ cm$^2$/Vs at $T = 300$ K which increases up to ~350 cm$^2$/vs at 2 K. Semiconducting behavior is observed at low electron densities $n$, but at high values of $n$ the resistivity decreases by a factor > 7 upon cooling to 2 K and displays a metallic $T^2$-dependence. This indicates that the band structure of 1T'-ReS$_2$ is quite susceptible to an electric field applied perpendicularly to the layers. The electric-field induced metallic state observed in transition metal dichalcogenides was recently claimed to result from a percolation type of transition. Instead, through a scaling analysis of the conductivity as a function of $T$ and $n$, we find that the metallic state of ReS$_2$ results from a second-order metal to insulator transition driven by electronic correlations. This gate-induced metallic state offers an alternative to phase engineering for producing ohmic contacts and metallic interconnects in devices based on transition metal dichalcogenides.

Transition metal dichalcogenides (TMDs) display band gaps ranging from ~1 to 4 eV and show promise in electronic and optoelectronic applications.[1,2] Several of these compounds were found to display extraordinary performance as field-effect transistors (FETs)[1-4] with high ON/OFF ratios[4-7] up to ~$10^8$, in addition to being flexible and nearly transparent[8] when compared to their bulk counterparts. Among the semiconducting TMDs, $MoS_2$,[5-9] $WS_2$,[10-14] and $WSe_2$,[15] are at the forefront of research because they are easy to synthesize, are naturally abundant, and display band gaps which typically fall within the visible range of light. Their indirect band-gap is comparable to that of Si, yet increases and becomes direct in monolayers.[16]

In contrast to $MoS_2$, $WS_2$, and $WSe_2$, $ReS_2$ is characterized by a distorted 1T′-structure with octahedral coordination that results in many unique properties. For example, the conductivity is anisotropic with larger carrier mobilities (by a factor of 3) for currents flowing along the *b*-axis, when compared to currents along the *a*-axis.[17] This crystallographic and concomitant electrical anisotropy were demonstrated to be convenient for fabricating digital inverters.[17] Furthermore, and in contrast with $MoS_2$,[16] $ReS_2$ displays a single direct band gap of 1.5 eV that is nearly independent on the number of atomic layers.[18] Recently, compounds displaying the 1T' structure, such as $WTe_2$ and $ReS_2$, were claimed to be susceptible to an intrinsic band inversion between chalcogenide *p*- and metallic *d*- bands.[19] Tuning the gap with an external electric field is predicted to lead to a phase-transition between a trivial topological state and state characterized by $Z_2 = 1$ topological invariant with concomitant quantum spin Hall-effect.[19]

Here, we report on the anisotropic Raman response of ReS$_2$ both experimentally and based on density-functional-theory calculations. In contrast to previous reports,[18] we indexed *all* peaks observed in its Raman spectrum as $A_g$-type first-order modes. We also studied the temperature-dependent four-terminal electrical transport properties of ReS$_2$, including field-effect mobilities (μ$_{FE}$), by studying FETs fabricated on few-layered samples exfoliated onto SiO$_2$/Si substrates. Our single-gated samples display a remarkable increase in the field-effect mobilities with decreasing temperature (*T*): the maximum μ$_{FE}$ increases from ~ 20 to ~35 cm$^2$/Vs at *T* = 300 K to ~ 350 cm$^2$/Vs at 2 K. Most importantly, for carrier densities (*n*) approaching $10^{13}$ cm$^{-2}$, all samples studied displayed a *metallic* resistivity ($\rho$) over the *entire T*-range, with $\rho$ increasing by nearly one order of magnitude upon cooling. This contrasts with the thermally activated behavior observed at lower densities. Through a scaling analysis of the conductivity as a function of the *T* and *n*, we demonstrate that this metal-insulator transition is not of percolation type,[20] but instead is quantum-critical like as observed in heterostructures of conventional semiconductors.[21-24] Hence, TMDs and in particular ReS$_2$, can open a new era for a long standing yet not fully understood problem:[24] the nature of the metal-insulator transition in two-dimensions that results from the interplay between disorder and electronic correlations. Our scaling analysis implies that this transition in ReS$_2$ belongs to the same universality class as the quantum metal to insulator transition established for silicon MOSFETs.[23] This simple but high degree of control on the electronic response of ReS$_2$ has a high potential for technological applications.

ReS$_2$ single-crystals were grown *via* the chemical vapour-transport (CVT) technique as described in the Supporting Information. Raman scattering, energy dispersive analysis (EDS), photoluminescence (PL) spectroscopy, schematics, optical images, and atomic force microscopy (AFM) images of our FETs, can be found in the supplementary information. The chemical and structural analysis indicates that the CVT process yields high purity ReS$_2$ crystals. Fabrication of our FETs is also described in the Supporting Information.

Figure 1 shows the Raman spectra of a few-layered ReS$_2$ crystal under different polarization conditions. We analyzed the vibrations of ReS$_2$, which belongs to the P-1 space group, by considering that its unit cell exhibits the symmetry of the C$_i$ point group in which the monolayer is centrosymmetric (has inversion symmetry). Using first principles calculations (see Methods), we found that there are 18 active Raman modes and 15 active infrared modes. The 18 Raman modes are all first order modes of $A_g$ type (see Fig. 1) whereas the infrared modes are of $A_u$ type. From our calculations, there is one in-plane mode in which all vibrations are parallel to the plane around 155 cm$^{-1}$ where both the Re and S atoms vibrate (labeled as $A_g^1$). In addition, there is one out-of-plane mode around 437 cm$^{-1}$ in which just the S atoms vibrate (labeled as $A_g^2$) and one quasi out-of-plane mode at 418 cm$^{-1}$ where the S vibrations are at a slight angle with respect to the plane (labeled as $A_g^3$). For all of the other modes, the atoms vibrate with an angle (except for 90°) with respect to the plane due to the very low symmetry of ReS$_2$. We chose to label these remaining 15 Raman active modes in order of increasing frequency (from $A_g^4$ to $A_g^{18}$), as shown in Figure 1. It is noteworthy that the peak around 305 cm$^{-1}$ is a superposition of two peaks that are not distinguishable with any of these spectra. Previous reports on ReS$_2$ sought to emphasize the in-plane vibrational modes by labeling certain peaks as $E$ modes (or $E$-like) in order to compare the modes with the Raman spectra of MoS$_2$ and WS$_2$. Strictly speaking, the $E$ or $E_g$ irreducible representations do not exist in the character tables of the $C_i$ point group which is the point group associated to the symmorphic space group P-1 of ReS$_2$: the only irreducible representations found in this point group are $A_g$ and $A_u$. We believe that the proposed labeling in

Figure 1a for the Raman modes should be adopted by the community working on transition metal dichalcogenides in referencence to ReS$_2$ to ensure a correct and consistent labeling scheme.

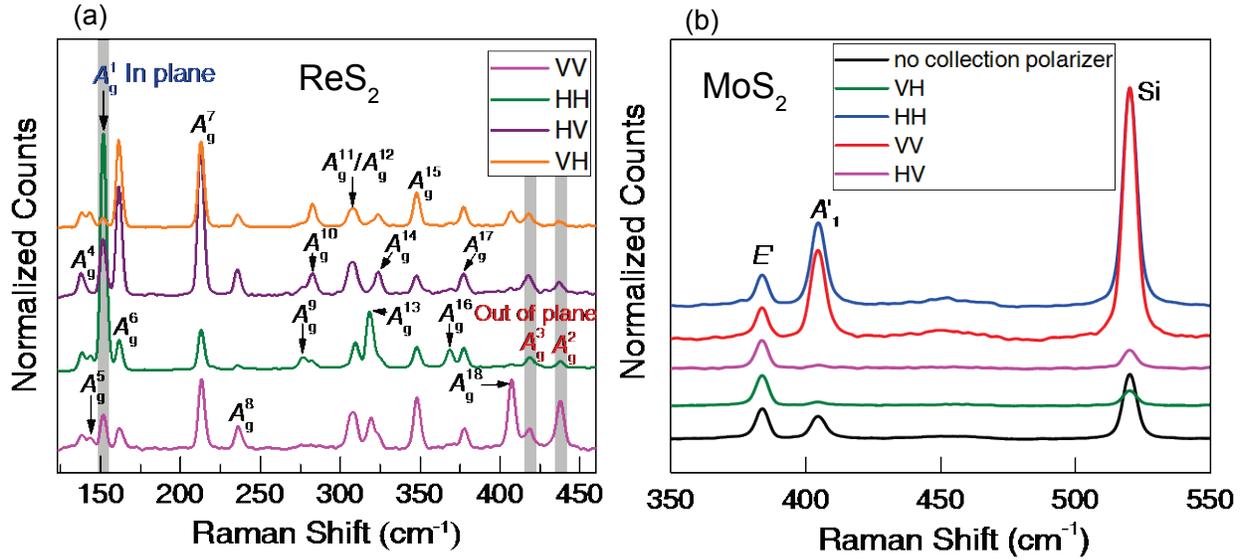

**Figure 1.** (a) Raman spectroscopy measurements performed on a few-layered ReS$_2$ flake with four different directions of light and collection polarizations. The spectra are normalized to $A_g^{17}$. The labeling of the peaks is based on density functional theory calculations. The most pronounced in-plane and out-of-plane modes are indicated by vertical gray bars. V and H stands for vertical and horizontal polarizations, respectively. (b) Raman spectra for a few-layered MoS$_2$ crystal and for distinct incident (and collection) polarizations. In contrast to ReS$_2$ the position of the Raman peaks are not polarization dependent.

The low symmetry of ReS$_2$ is further confirmed from the polarization dependence of the peak intensities. In this context, we studied different polarization configurations for ReS$_2$, including VV, HH, VH, and HV where in "AB," the A refers to the incident polarization configuration and B refers to the collection polarization configuration. In this experimental setup, V refers to the 'X' direction while H refers to the 'Y' direction when viewing the sample from the stage. This data is shown in Figure 1a. As seen in Figure 1a, the peak intensity ratios are extremely sensitive to the direction of polarization, with the in-plane $A_g^1$ mode becoming most intense for the HH configuration.

Furthermore, the peak intensity ratios are not the same for the VH and the HV configurations. If this is compared with the polarization dependence of monolayer $MoS_2$ (Figure 1b), which belongs to the $D_{3h}$ point group (*S3*), it is clear that VV = HH and HV = VH. The fact that for $ReS_2$, VV ≠ HH and VH ≠ HV demonstrates the importance of the direction of excitation with respect to the crystal axes in determining the dominating vibrational modes. It should also be mentioned that the intensities in all of these spectra will change when the crystal is rotated even with the same laser and collection polarization, demonstrating the significant anisotropy of the phonon modes in $ReS_2$.

From this point on, we discuss the electrical transport properties of FETs built from multi-layered $ReS_2$ single-crystals. In order to minimize the influence of the contacts and to extract the intrinsic transport properties, these FETs were measured in a four-terminal configuration. The data in this manuscript was collected on three samples labelled as sample #1 (~12 atomic layers (*18*)), sample #2 (~11 layers), and sample #3 (~3 layers). The dimensions of are $l/w$ = 6.16 μm/1.9 μm (sample #1), 5.59 μm/5.1 μm (sample #2), and 3.89 μm/3.18 μm (sample #3), respectively. $l$ and $w$ are the length and the width of the channel, respectively. Two-terminal measurements (from two other samples) are presented in the Supplementary Information to compare with the four-terminal measurements discussed below.

Figure 2 presents an overall evaluation of the transport properties of a mechanically exfoliated multilayered $ReS_2$ single-crystal (sample #1) upon which we thermally evaporated several contacts having different channel lengths (*l*). Simple schematics of our samples are shown in Figure 1a and an optical micrograph of sample #1 is shown in Figure 2b. Here the current is injected along the long axis of the crystal which corresponds to the *b*-axis. Through optical inspection and by analyzing the angles at the edges of the crystals we found that $ReS_2$ exfoliates as elongated platelets with the long axis being the *b*-axis. Figure 2c displays the room temperature drain-to-source current $I_{ds}$ as a function of the back-gate voltage $V_{bg}$ under a fixed bias voltage $V_{ds}$ = 150 mV, and for several values of *l*. A sizeable $I_{ds}$ can be extracted for $V_{bg}$ > -20 V indicating that the as-grown material is electron doped, due perhaps to S vacancies as claimed to be the case for $MoS_2$.[25, 26]

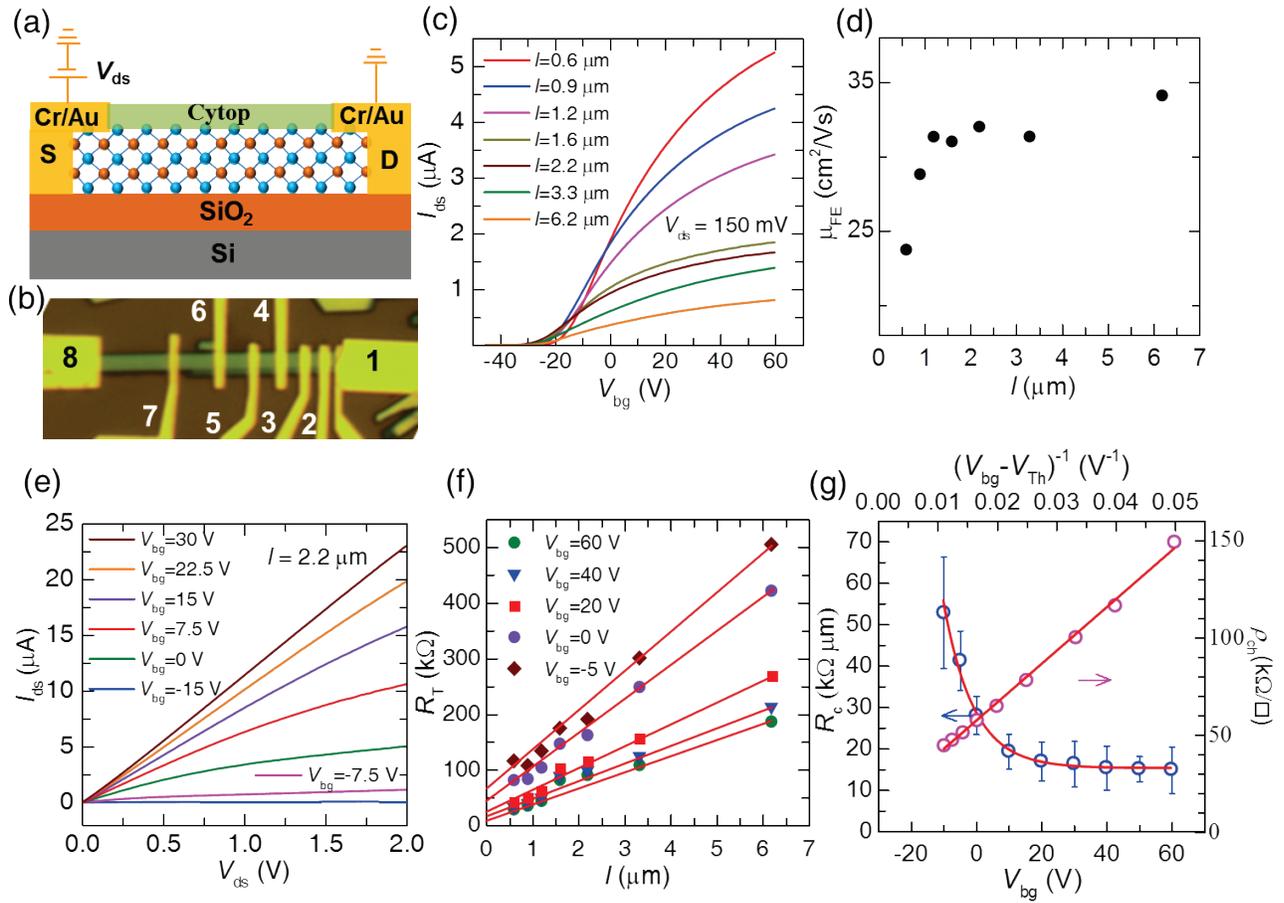

**Figure 2.** (a) Sketch of $ReS_2$ field-effect transistor on $Si/SiO_2$ substrate with 2-terminal and bottom gate configuration. (b) Optical micrograph of a $ReS_2$ single-crystal mechanically exfoliated onto $SiO_2$ showing contacts of distinct separation(s) intended for four-terminal and for transfer length method (TLM) measurements. The total length of the channel i.e. between contacts 1 and 8 is $l_c \cong 16$ μm and the width of the channel is $w$ = 1.6 μm. Atomic force microscopy measurements indicate that this crystal is $t$ = 8.1 nm thick, or composed of approximately 11 atomic layers. (c) Room temperature two-terminal transfer characteristics or drain to source current $I_{ds}$ as function of the back-gate voltage $V_{bg}$ for several distances *l* between the contacts. These curves were collected under a bias-voltage $V_{ds}$ = 150 mV. Notice that the transistor is in the off state for $V_{bg} \leq$ -30 V. (d) Maximum field-effect mobilities, as extracted from the linear slopes in the traces in (B), and as a function of *l*. (e) $I_{ds}$ as a function of $V_{ds}$ for several values of the back gate voltage $V_{bg}$, as collected among contacts and which have a separation *l* = 2.2 μm. Notice the nearly linear dependence of $I_{ds}$ on $V_{ds}$. (f) Two-terminal resistance $R_T$, which includes both the resistance of the contacts and that of the channel, as a function of the separation $l$ between contacts. Red lines are linear fits. (g) Contact resistance $R_c$ (left axis), as extracted from the linear fits in (f), as a function of $V_{bg}$. Red line is a fit to an exponential dependence. The same panel also presents the resistivity of the channel $\rho_c$ (right axis) as a function of $(V_{bg} - V_{Th})^{-1}$. Red line is a linear fit from which one evaluates the intrinsic mobility $\mu_i$.

Unsurprisingly, we find that the shorter the channel length $l$, or the shorter the channel resistance, and hence the higher the current extracted under a fixed bias voltage. Figure 1D displays the maximum room temperature, two-terminal field-effect mobilities, i.e., $\mu_{FE} = [dI_{ds}/dV_{bg}] \times [L/WCV_{ds}]$, where $C = \varepsilon_r \varepsilon_o/d$ is the gate capacitance per unit area, $\varepsilon_r \sim 3.9$ is the dielectric constant of $SiO_2$, $\varepsilon_o$ is the dielectric constant of free space, and $d \cong 285$ nm is the thickness of the dielectric layer. Therefore, the corresponding geometrical gate capacitance is $C = 12.783 \times 10^{-9}$ F/cm$^2$. Here, $dI_{ds}/dV_{bg}$ was obtained from a simple linear fit of $I_{ds}(V_{bg})$ in the region where $I_{ds}$ increases linearly with $V_{bg}$. The two-terminal field-effect mobility is observed to increase quickly and to saturate as soon as $l > 1$ at values ranging between 30 and 35 cm$^2$/Vs. Such an increase in mobility as function of $l$ was reported by several groups in graphene see, for example Ref.[27], and attributed either to i) the deterioration of the material due to e-beam irradiation around the contact area or to ii) a crossover from a diffusive to a ballistic transport regime. The considerably smaller mobilities of ReS$_2$ when compared to graphene, seems inconsistent with a ballistic regime, therefore the short channel effect on $\mu_{FE}$ is most likely associated with the deterioration of the material upon irradiation. The nearly independence of $\mu_{FE}$ on $l$, indicates that the crystal is rather homogeneous and that $\mu_{FE}$ is limited by phonons and impurity scattering. Figure 2e displays $I_{ds}$ as a function of $V_{ds}$, measured in a two-terminal configuration through contacts which are placed $l = 2.2$ μm apart. The $I$-$V$ characteristics are linear particularly for $V_{ds} < 1$ V, indicating that thermionic emission processes promote charge carriers across the Schottky barriers, which is given by the difference between the electron affinity of ReS$_2$ and the work function of Cr. It also suggests that the barriers are narrow, i.e., the probability of carrier tunnelling through them is high. Figure 1F displays the total two terminal resistance $R_T = 2R_c + \rho_{ch} l/w$, where $R_c$, $\rho_{ch}$ and $w$ are the resistance of the contacts, the resistivity of the channel and its width, respectively, as a function of the channel length $l$ and for several values of the gate voltage $V_{bg}$. Red lines are linear fits from whose intercept and slope we extract $R_c$ and $\rho_{ch}/w$, respectively. The slope or $\rho_{ch}$, decreases as $V_{bg}$ increases due to the progressive accumulation of electrons into the channel. Figure 1G displays $R_c$ as a function of $V_{bg}$ and $\rho_{ch}$ as a function of $(V_{bg}-V_{th})^{-1}$ where $V_{th}$ is the threshold gate voltage for carrier conduction, i.e. $V_{th} \cong -30$ V. $R_c \cong 53$ k$\Omega$ µm at $V_{bg} = -10$ V decreasing exponentially (red line is an exponential fit) as $V_{bg}$ increases and subsequently saturating at $R_c \cong 15$ k$\Omega$ µm. The initial value is ~2 times smaller while the later saturating value is ~3 times higher than those reported by Ref.[28] indicating similar quality of contacts. Given that $\rho_{ch} = \sigma_{ch}^{-1} = (ne\mu_i)^{-1}$, where the total charge accumulated in the channel is given by $ne = (V_{bg} - V_{th})C$ it is possible to evaluate the intrinsic carrier mobility $\mu_i$ through the slope of $\rho_{ch}$ as a function of $(V_{bg} - V_{th})^{-1}$. We obtain $\mu_i \cong 29.5$ cm$^2$/Vs which, as we will show below is comparable to the four-terminal field-effect mobilities extracted by us for ReS$_2$ at room temperature.

Figures 3a and 3b display $I_{ds}$ of sample #1 in a logarithmic scale as a function of $V_{bg}$, under a constant bias voltage $V_{ds} = 100$ mV, and for several temperatures $T$ using either a two-terminal (contacts 1 and 4) or a four-terminal (contacts 1, 2, 3, 4) configuration, respectively. As seen, the ON/OFF ratio ranges from $10^4$ to $10^5$ which is comparable to earlier reports on monolayer or few-layered MoS$_2$, MoSe$_2$ and WSe$_2$.[3,4,6] Several ReS$_2$ based FETs were characterized and all of them showed similar *n*-type behavior which is consistent with recent reports on ReS$_2$ FETs using similar metals for the contacts.[17, 27] The four-terminal temperature dependent data for sample #3 ($T$ = 300 K down to 2 K) and sample #2 ($T$ = 340 K down to 2 K) are displayed in Figure 3c and in Supplementary Figure S2. The threshold gate voltage varies slightly from sample to sample. At room temperature for sample #1 under $V_{bg}$ = 0 V, a charge carrier density $n \sim 6.7 \times 10^{12}$/cm$^2$ is estimated through $n = \frac{C}{e}|V_{Th} - V_{bg}|$. For sample #2, a maximum room temperature field-effect mobility of $\mu_{FE} \cong 19$ cm$^2$/Vs is obtained from the slope of $I_{ds}$ as a function of $V_{bg}$. All curves were measured under a fixed excitation voltage $V_{ds}$ = 150 mV. Although this value is similar to recently reported room temperature mobilities for ReS$_2$ FETs,[28] it is nevertheless much lower than the room temperature mobilities reported for few-layered $n$-MoS$_2$, $p$-WSe$_2$ and ambipolar MoSe$_2$ FETs[4,6,15] suggesting a far more pronounced role for phonon scattering or heavier effective masses. Notice how the slope of $I_{ds}$ as a function of $V_{bg}$ increases as the temperature decreases, indicating an increase in carrier mobility. The threshold gate-voltage is also observed to increase in all samples, e.g. in sample #2 it is displaced from ~ -35 V to ~+10 V with decreasing temperature, which can be ascribed to charge localization at the interface between the semiconductor and the substrate.[15] Localization in the channel would result from disorder, e.g. substrate roughness or randomly distributed charges on the SiO$_2$ layer.[18]

The extracted field-effect mobility as a function of the temperature for all three devices is plotted in Figure 3d. All three samples show a similar mobility trend; $\mu_{FE}$ increases considerably as the temperature is lowered. $\mu_{FE} \sim 350$ cm$^2$/Vs is the maximum four-terminal field-effect mobility (extracted from sample #2) observed at $T$ = 2 K. The four-terminal $\mu_{FE}$ was evaluated from the MOSFET transconductance formula by replacing the bias voltage $V_{ds}$ with the voltage $V_{23}$ across the voltage leads 2 and 3 (see Figure 1b). Our obtained values are considerably higher than those recently reported for the same compound at 77 K.[28] Such a marked difference in the low temperature field-effect mobilities is most likely attributable to the role of contacts. For example, notice how the two-terminal mobility as extracted from sample #1 (solid blue markers) decreases to a value of 25 cm²/Vs, which is considerably smaller than the values extracted from both sample #1 and sample #2 when using a four-terminal configuration at similar temperatures.

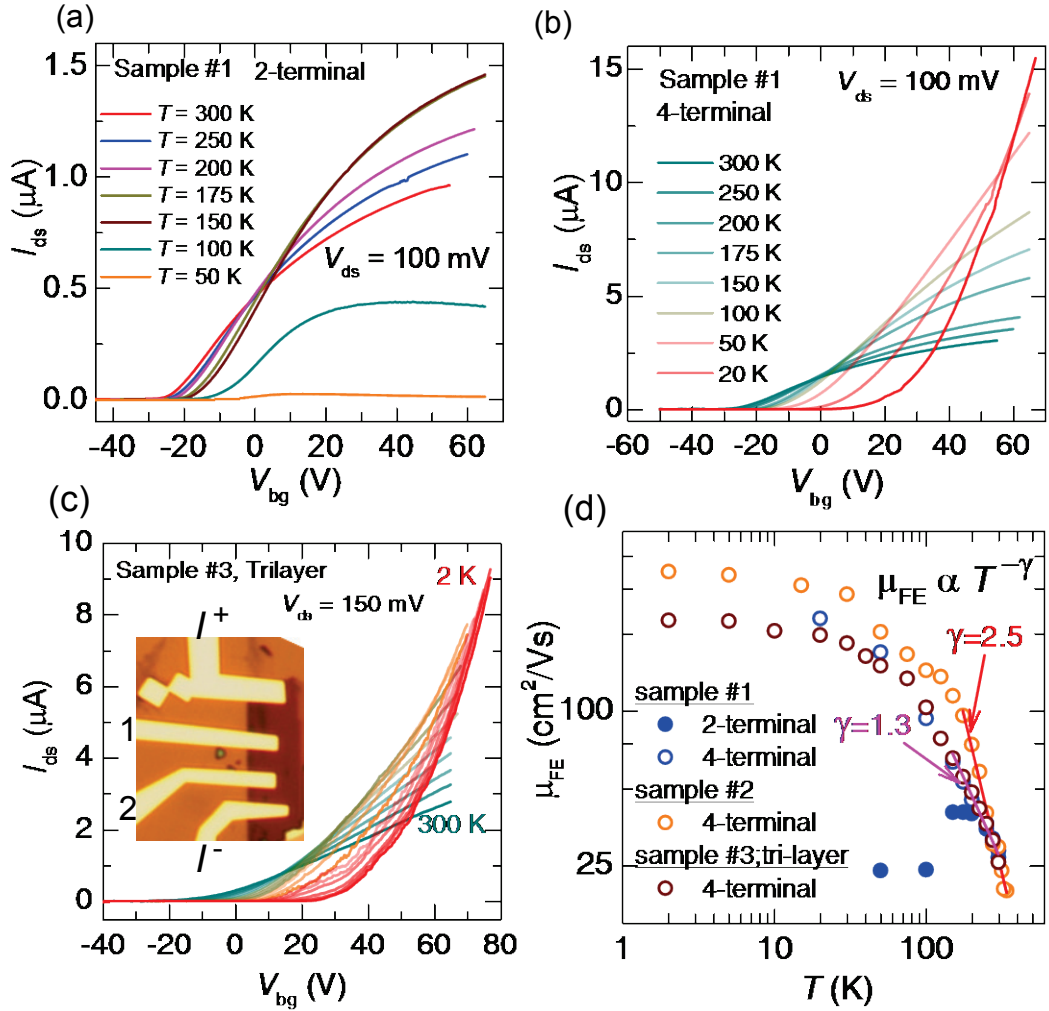

**Figure 3.** (a) Logarithmic plot of $I_{ds}$ as a function of $V_{bg}$, as measured *via* a two-terminal configuration or through contacts 1 and 4 of sample #1 for several temperatures ranging from $T$ = 300 K down to 50 K under a fixed bias $V_{ds}$ = 100 mV. (b) Logarithmic plot of $I_{ds}$ as a function of $V_{bg}$, as measured *via* a four-terminal configuration or through contacts 1 and 4 (current) and 2 and 3 (voltage) from sample #1 and for several temperatures ranging from $T$ = 300 K down to 20 K under a fixed bias $V_{ds}$ = 100 mV. (c) Same as in (b) but for sample #3 (tri-layered crystal) measured in a four-terminal configuration from 300 K to 2 K. Inset: 4-terminal optical image of the device with Cr/Au contacts. $I^+$, $I^-$ depicts the current contacts and 1 and 2 depicts the voltage contacts used for 4-terminal measurements. (d) Log-log plot of the maximum field-effect mobility $\mu_{FE}$ as a function of the temperature for sample #1, as measured through a four-terminal (open blue makers) and a two-terminal configuration (solid blue makers). Open orange

markers depict four-terminal field-effect mobilities for sample #2 while the brown ones correspond to four-terminal field-effect mobilities extracted from sample #3, which is composed of three atomic layers. Solid lines are linear fits yielding sample-dependent exponents γ in $\mu_{FE} \propto T^{-\gamma}$ ranging between 1.3 (3 layers) and 2.5 (~ 10 layers).

We measured several samples with thicknesses ranging from 6 to 12 atomic layers and obtained similar values for the two-terminal mobilities.

Earlier studies on WSe$_2$, and on MoS$_2$ indicates that carrier mobilities tend display a maximum for samples composed of ~8 to 12 layers.[15,29] Our samples display a strongly stemperature dependent mobility, indicating that it is limited by carrier scattering from phonons, with the scattering rate decreasing as the temperature is lowered. The sharp increase in mobility occurs mainly between room temperature and 100 K. The saturation of the mobility below 100 K suggests that impurity scattering, carrier localization, or the suppression of thermionic emission of carriers across the Schottky barriers limit charge transport at low temperatures.[30-32] We fitted the high temperature mobility to a power law, *i.e.* $\mu_{FE} \propto T^{-\gamma}$, where the exponent γ is given by the phonon scattering mechanism. From these fits, we extracted γ = 1.6 for sample #3 and γ = 2.5 for sample #2, which is comparable to the exponents extracted for the electron mobilities of Ge (γ=1.7) and Si (γ=2.4) and to theoretical predictions concerning *n*-type MoS$_2$ FETs.[30] Undoubtedly, this strong *T*-dependence indicates that phonons are powerful scatterers in ReS$_2$. The pronounced sample-dependence also indicates that strain, impurities and the interaction with the substrate affect the value of γ.

In Figure 4, we display and analyze the temperature dependence of the four-terminal conductivity σ$_{4T}$. Figure 4a shows σ$_{4T}$ in a logarithmic scale and as function of the inverse temperature $T^{-1}$. At negative gate voltages, when the FET is nearly in the OFF state it displays semiconducting behavior with a small and gate-dependent gap ranging from 1000 to 100 K, as extracted from the linear fits (red lines). Deep in the OFF state ($V_{bg} < -20$ V), the sample becomes extremely insulating at lower temperatures and our experimental set-up prevents us from extracting the intrinsic semiconducting gap of ReS$_2$. Clearly, the application of a gate voltage pins the Fermi level at an arbitrary position, likely at an impurity level between the valence and the conduction bands, thus leading to smaller activation gaps. This is illustrated by Figure 4b, which shows the $\rho_{4T} = \sigma_{4T}^{-1}$ in a logarithmic scale as a function of the inverse temperature $T^{-1}$. The red line is a linear fit from which we extract an activation energy $\Delta = 0.12$ eV, which is considerably smaller than the optical gap of ~ 1.5 eV of ReS$_2$ (see Supplementary Figure S1 for the PL data). At lower $T$s we found that $\sigma_{4T}(T)$ crosses-over to a two-dimensional (2D) variable-range hopping conductivity. In contrast and as seen in Figure 4c, at higher gate voltages, $\sigma_{4T}$ increases over the entire range of temperatures as $T$ is lowered, indicating metallic conductivity. To more clearly expose this metallic dependence, Figure 4d displays the four-terminal resistivity $\rho_{4T} = \sigma_{4T}^{-1}$ as a function of $T$. As seen, between room temperature and 2 K, $\rho_{4T}$ decreases by a factor greater than 7. The dashed red line is a fit to a metallic Fermi liquid like $T$-dependence, which describes $\rho_{4T}(T)$ remarkably well over the *entire T*-range. In Figure 5, we show through a scaling analysis, that one can collapse all the $T$-dependent conductivity curves into two branches, i.e. a metallic and an insulating one. This indicates that a second-order metal to insulator transition occurs in this system as a function of carrier density, as illustrated by the $T$-dependent data in Figures 4a to 4d. A metal-insulator transition is claimed to occur in MoS$_2$ FETs using a high-κ dielectric such as HfO$_2$, in combination with a top gate to accumulate a high carrier density ( >$10^{13}$ cm$^{-2}$) at the interface with the semiconductor.[33] The initial reports on the metal-insulator transition in 2D electronic systems,[21,22] claimed that the metallic behavior results from the interplay between disorder or randomness and strong electronic correlations.[33] In effect, according to the scaling theory of localization,[35] for 2D systems of non- or weakly-interacting electronic systems, disorder and randomness would always cause the conductivity to decrease as the temperature is lowered, leading to an insulating ground-state.

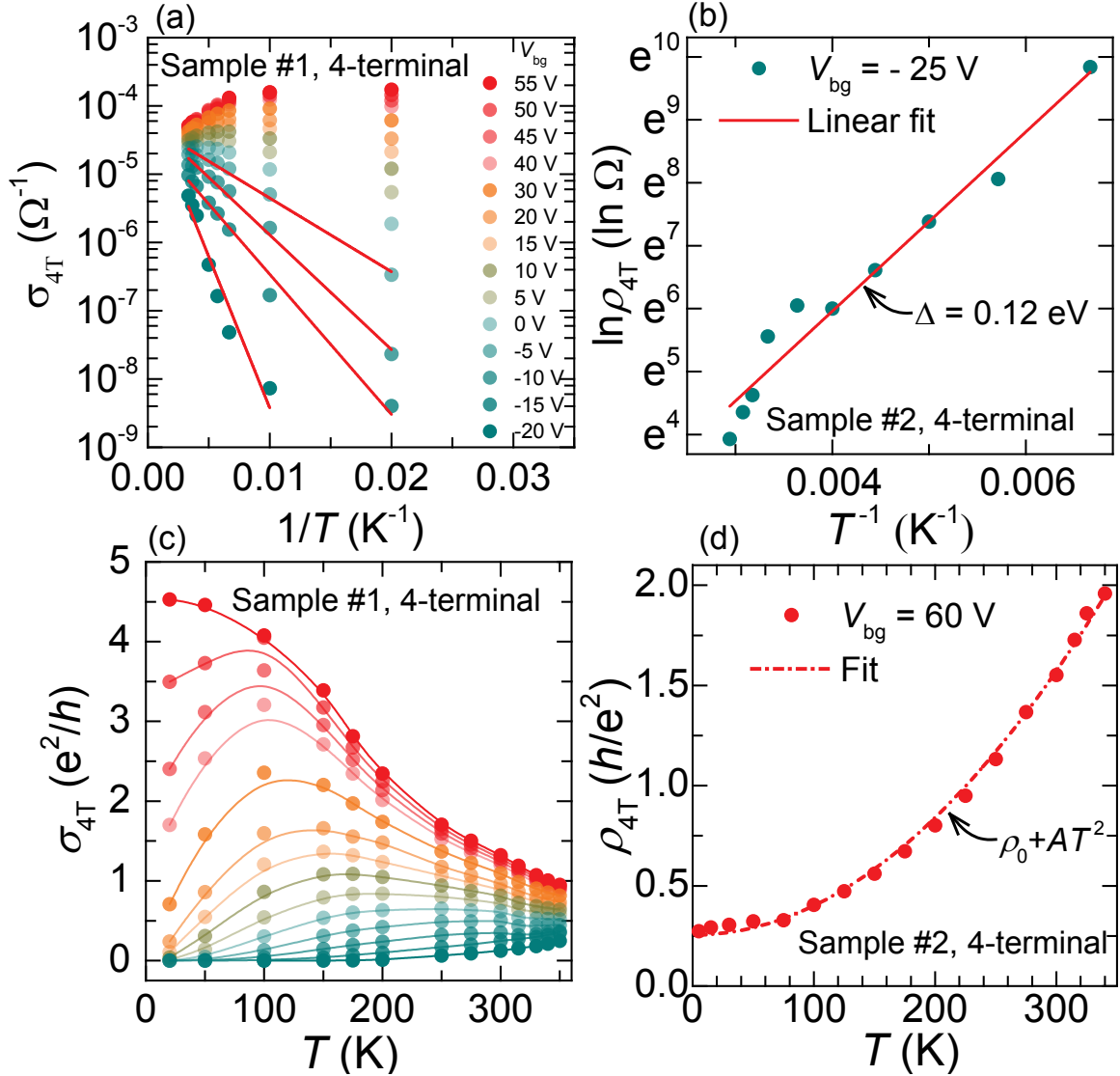

**Figure 4.** (a) Four-terminal electrical conductivity $\sigma_{4T} = \left(\frac{I_{ds}}{V_{ds}}\right)\left(\frac{l}{w}\right)$, where $l$ is the separation between the voltage contacts 2 and 3 and, $w$ is the width of sample #1, as a function of $1/T$ under several values of $V_{bg}$. For the points collected at lower gate voltages, red lines are linear fits indicating activated behavior with activation energies ranging from ~ 0.038 eV under $V_{bg}$ = -20 V to ~0.013 eV under $V_{bg}$ = - 5 V . (b) Four-terminal resistivity $\rho_{4T} = 1/\sigma_{4T}$ from sample #2, in a natural logarithmic scale as a function of the inverse temperature and as collected under a gate voltage $V_{bg}$ = - 25 V. Red line is linear fit from which we extract an activation energy $\Delta$ = 0.12 eV. (c) $\sigma_{4T}$ in units of $(e^2/h)$ as a function of the temperature. In contrast to what is observed under $V_{bg}$ = - 10 V for which $\sigma_{4T} \rightarrow 0$ upon cooling, $\sigma_{4T}$ increases by a factor > 3 under $V_{bg}$ = 55 V. (d) $\rho_{4T}$ from sample #2 measured under $V_{bg}$ = 60 V and as function of T. As seen, $\rho_{4T}$ decreases by a factor > 7 upon cooling. Dotted line is a fit to $\rho_{4T} = \rho_0 + AT^2$, indicating that Fermi liquid like metallic behavior over the entire T range.

For example, when carriers are scattered by impurities back to their initial trajectory, their electronic waves interfere constructively with their time reversed paths, leading to electron localization, or weak localization particularly at low temperatures.[36] Nevertheless, the scaling theory does not explicitly consider the effect of the Coulomb interaction between carriers whose strength is characterized by the dimensionless Wigner-Seitz radius:

$$r_s = \frac{n_v}{a_B^* \sqrt{\pi n_{2D}}} = \frac{m^* e^2 n_v}{4\pi\varepsilon\hbar^2 \sqrt{\pi n_{2D}}} \qquad (1)$$

Here, $n_v$ corresponds to the number of valleys, $m^*$ to the effective mass, $\varepsilon$ to the dielectric constant of the material, $n_{2D}$ to the density of the 2D electron gas, $e$ to the electron charge, and $\hbar$ to the Planck constant. For $r_s \ll 1$, all the published experimental evidence supports the scaling theory of localization. However, for the strongly interacting limit, i.e. $r_s \gg 1$, several authors[34,37] predicted that for weak disorder, a 2D system should scale towards a conducting state as the temperature is lowered. This is supported by experiments on low-disordered 2D silicon samples [20-22] demonstrated that for strongly interacting systems, or for $r_s > 10$, one can cross from an insulating to a metallic regime with increasing electron density. For monolayer MoS$_2$, $r_s$ is estimated to be $\approx$ 4.2 (similar to the $r_s$ value for GaAs/AlGaAs heterostructures) at the critical density $n_{2D} = n_c \sim 1 \times 10^{13}$ cm$^{-2}$ where the insulator to metal transition is claimed to occur, thus suggesting that a MoS$_2$ monolayer is characterized by strong Coulomb correlations.[33] In ReS$_2$, the low temperature mobilities in Figure 3d are comparable, or somewhat smaller, than those reported for monolayer or bilayer MoS$_2$,[32] suggesting a similar or a comparatively higher effective mass. The scaling analysis shown below and also in the supplementary information, in combination with Hall-effect measurements (see Supplementary Figure S3), would indicate sample-dependent $n_c$ values ranging from 2 to 5 x $10^{12}$ cm$^{-2}$. To our knowledge $\varepsilon$ for ReS$_2$ has yet to be reported, but if we assume it to be within a factor of 2 to 3 of the value reported for MoS$_s$ ($\varepsilon=7.3\varepsilon_0$) (32), $r_s \gg 1$ would also be satisfied. In Si MOSFETS, the evidence for a second-order metal-insulator transition and its associated quantum critical point is provided by the scaling of the conductivity $\sigma$ as a function of the carrier density $n_s$ at low temperatures.[21-23, 37, 38] Here, low $T$s correspond to $k_BT \ll k_BT_F = E_F = \pi\hbar^2 n_s/m^*$, which for $m^* = 0.5m_0$ and for $10^{12}$ cm$^{-2} \leq E_F \leq 10^{13}$ cm$^{-2}$ would correspond to ~56 K $\leq T_F \leq$ 556 K.

According to a scaling theory[38] of the metal insulator transition in 2D, the temperature dependence of the conductivity in the critical region is given to leading order by:

$$\sigma(\delta_n, T) = \sigma_c(T) F[T/T_0(\delta n)]) \qquad (2)$$

where $\sigma_c(T \to 0 \text{ K}) \propto T^x$ is the conductance at the critical density $n_c$, $\delta n \equiv (n_s - n_c)/n_c$, $z$ is the dynamical exponent, and $\nu$ is the correlation length exponent.[38] $T_0$ is a crossover temperature corresponding to the inverse of the correlation time as $T_0(\delta n) \sim |\delta n|^{\nu z}$.

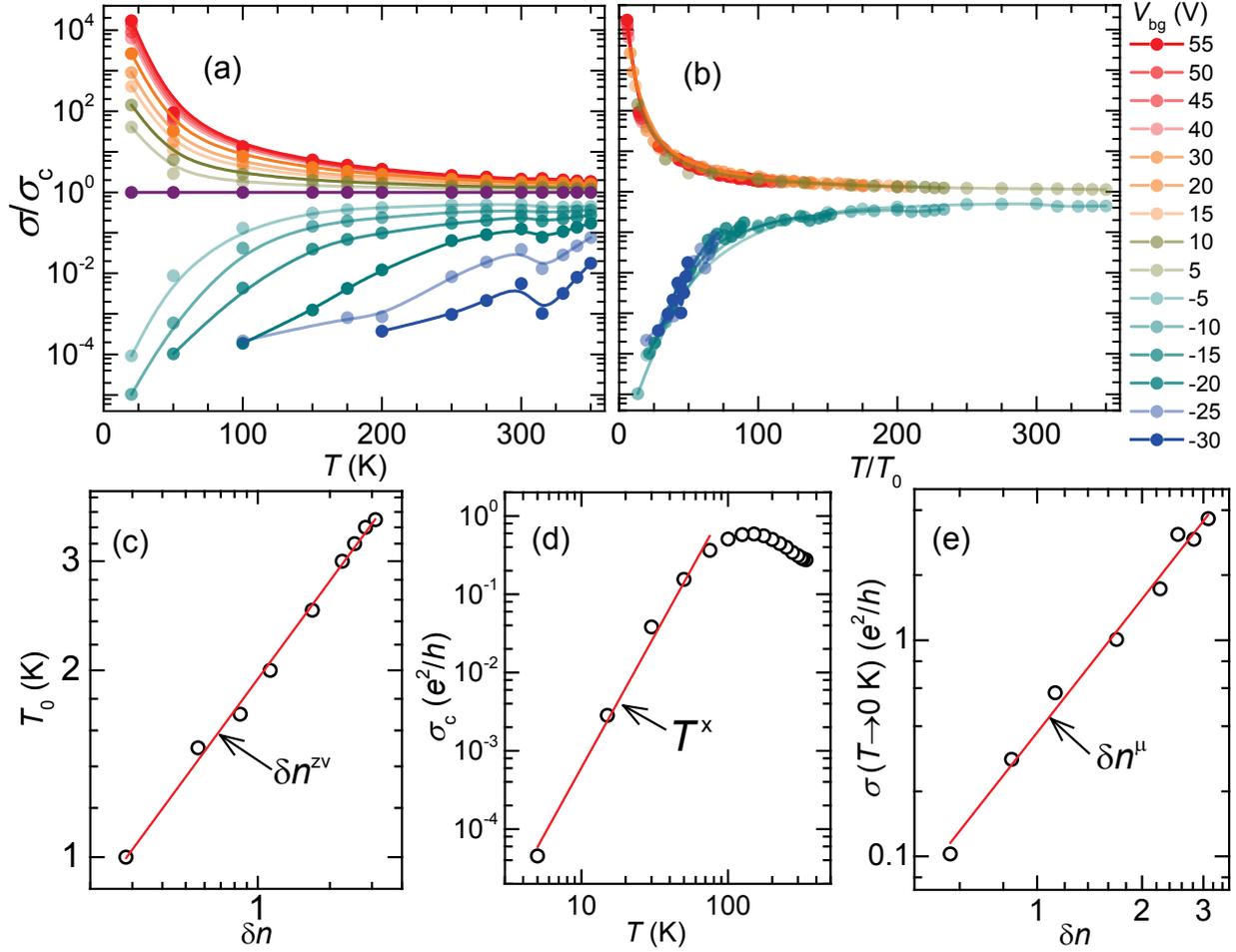

**Figure 5.** (a) Conductivity $\sigma$ curves renormalized by the conductivity $\sigma_c(T, V_{bg} = 0\ V)$ (depicted by purple dots) at the critical density $n_c = n(V_{bg} \cong 0\ V)$ as functions of the temperature $T$. (b) Re-scaled conductivity $\sigma/\sigma_c$ as a function of the re-scaled temperature $T/T_0$ yielding metallic and insulating branches, respectively. (c) Renormalization temperature $T_0$ as a function of $\delta n = (n_H - n_c)/n_c$ where $n_H$ is the carrier density extracted from the Hall-effect. Red line is a linear fit, yielding the critical exponent $z\nu = (0.6 \pm 0.1)$. (d) $\sigma_c$ as a function of $T$ in a log-log scale. Red line is a linear fit, which yields the critical exponent $x = (3.4 \pm 0.1)$. (e) Zero-temperature conductivity $\sigma(T = 0\ K)$, obtained from a extrapolation of $\sigma(T \rightarrow 0\ K)$ as a function of $\delta n$. Red line is a linear fit, which yields the critical exponent $\mu = (2.1 \pm 0.1)$, thus satisfying $z\nu x \cong \mu$.

Therefore, the temperature dependence of $\sigma$ is predicted to be exponential both on the insulating and on the metallic sides of the transition, and with the same exponent. The results of this scaling for the raw data in Figure 4c are shown in Figure 5. As seen through Figures 5a and particularly in 5b, all the data collapses into two branches: the upper one for the metallic side of the transition and the lower one for the insulating side. As indicated in Figure 5c, the best collapse is found for $z\nu = (0.6 \pm 0.1)$, over a remarkably wide range of electron densities (with $\delta n$ all the way up to 3). The $n_s(V_{bg})$ was determined through Hall-effect measurements performed on the same sample (see Figure S3). Hall-effect measurements indicate very small, fast vanishing carrier densities for increasing *negative* values of $V_{bg}$. Hence, we could not reliably extract $z\nu$ for $\delta n < 0$.

$\sigma_c$ was chosen as the $\sigma(n_s,T)$ trace (in Figure 5c) which did not display a clear activated (or insulating) nor an obvious metallic behavior at low $T$s. Instead, as illustrated by Figure 5d, we chose $\sigma_c(n_c,T)$ as the trace displaying an anomalous power-law dependence on $T$, i.e. $\sigma_c(n_c,T) \propto T^x$, over more than a decade in temperature at low $T$s. In this case, we extracted an exponent of $x = (3.4 \pm 0.1)$ for $n_c \cong 3.02 \times 10^{12}$ cm$^{-2}$, which is the value extracted under $V_{bg} = 0$ V. It turns out that from standard scaling arguments,[34] the critical exponent $\mu$ relating $\delta n$ to the zero-temperature conductivity (determined from the extrapolations of $\sigma(n,T)$ to $T = 0$ K), or $\sigma(\delta n, T=0$ K$) \propto T^\mu$, is also related to the other critical exponents through the relation $\mu = x(z\nu)$. Indeed, using $x \cong (3.4 \pm 0.1)$ (Figure 5d) and $z\nu \cong (0.6 \pm 0.1)$ (Figure 5c), one obtains $\mu = x(z\nu) = (2.0 \pm 0.2)$ which is in remarkable agreement with $\mu = (2.1 \pm 0.1)$ extracted from the $T \to 0$ K extrapolation of $\sigma(n,T)$ as a function of $\delta n$ (Figure 5e). This provides consistency to our scaling analysis. Notice that Ref.[39] also claims, through two-terminal measurements, to observe a metal insulator transition in ReS$_2$, although the claimed metallic phase is confined to temperatures above ~ 150 K, which precludes a detailed scaling analysis as presented here. The observed field-effect mobility is observed to saturate at a value of 1 cm$^2$/Vs below 10 K[39], which contrasts markedly with $\mu \cong 350$ cm$^2$/Vs observed by us for the 4-terminal field-effect mobility at low temperatures.

Recent theoretical work based on the holographic duality strongly suggests the possibility of nontrivial values for $x$ and also for $\mu$, which would be consistent with the existence of a sharp MIT in a correlated 2D electronic systems.[40] However, our exponents should be taken with a caution since critical exponents are notoriously difficult to extract. In the present case, we will need to approach the quantum critical point by performing detailed measurements down to much lower temperatures and over a broad range to carrier densities. Such a detailed study will be the subject of a future report. In the supplementary information we have included the scaling analysis for sample #3 (see Figure S4), which yields distinct critical exponents when $n$ is estimated through $n = e^{-1} C_{bg} |V_{bg} - V^t_{bg}|$, where $C_{bg}$ is the back-gate capacitance. For example, in this case one obtains $zv = (1.3 \pm 0.1)$ which is more in line with the values reported for disordered Si MOSFETS.[22,23]

A clarification is needed concerning the overall temperature dependence of $\sigma(n, T)$ shown in Fig. 3c. As seen, $\sigma(n, T)$ tends to display metallic behavior at higher $T$s, and reaches a maximum at a certain temperature $T^*(n)$, which is followed by insulating behavior upon subsequent cooling. In Si MOSFETS, the application of a substrate bias changes the electric field at the Si-SiO$_2$ interface. Consequently, the average position of the 2D electrons relative to the interface can change as well as the splitting between the subbands. Although at low temperatures all electrons are expected to populate only the lowest subband, at higher temperatures electrons could be thermally excited to band tails associated with the upper subbands which are inherent to the interfacial triangular quantum well. This upper subbands might be characterized by distinct mobilities and/or concomitant and effective masses, thus contributing to $\mu(T)$. Figure 4c clearly shows that $T^*$ is $n$-dependent which, is consistent with this simple scenario. More importantly, the renormalization of all $\sigma(n,T)$ curves by $\sigma_c$ accounts for this effect by renormalizing the $T$- dependent background associated with it, and yields in this way the intrinsic critical behavior of $ReS_2$ shown in Figure 5b.

Recently, the metal-insulator transition observed in monolayer and multilayer molybdenum disulphide FETs was claimed to be a percolation-type of metal-insulator transition, driven by density inhomogeneities of electronic states.[20] But it is important to emphasize that the power-law type of scaling performed here is *inconsistent* with a percolation like transition,[38] and is fully consistent with a quantum phase transition. The existence of a quantum metal-insulator phase-transition in a relatively disordered system characterized by high Fermi energies such as $ReS_2$, which has structural and physical properties so distinct from those of conventional semiconductors, opens up new possibilities for the field of quantum-criticality in Mott-Anderson systems.

It also opens up interesting technological possibilities, such as relatively simple to fabricate optoelectronic switches, i.e. whose electrical conduction or optical absorbance is controllable through a back gate, and without complex architectures involving multiple gates or a high-κ dielectric layer. More importantly, it also indicates that one can use the same material for both the semiconducting channel and for the metallic interconnects. These would be generated by a patterned grid acting as the electrical gate and insulated from the two-dimensional channel by a dielectric layer. In this way, band mismatch or Schottky barriers at the level of the electrical contacts, could be circumvented which insofar have been one of the greatest obstacles for the development of optoelectronics based on transition metal dichalcogenides. This is a simpler approach than the so-called phase engineering process which requires either a chemically[41]- or a heat-induced[42] phase-transformation from semiconducting $2H$ to semi-metallic $1T$-phase upon which the electrical contacts are deposited. The first approach leads to a metastable and inhomogeneous 1*T* metallic phase. The second vaporizes most of the material and induces a large amount of vacancies in the remnant layers. Both approaches lead to a structural mismatch in contrast to the approach proposed here which offers a far greater level of control on the electronic and structural properties of the active transition-metal dichalcogenide atomic layer(s) used for optoelectronic applications.

Finally, and as suggested by Ref.[19], the remarkable tunability of the ground state of $ReS_2$ through the application of an electric field (perpendicularly to the atomic layers), may open entire new avenues of research if band inversion was confirmed for a certain range of carrier densities.

**ASSOCIATED CONTENT** Supporting Information
The Supporting Information is available free of charge on the ACS Publications website at.

Methods: crystal Synthesis, device Fabrication, Raman, photoluminescence and transport measurements, Raman calculations, Figure S1; schematics and optical micrograph for sample #2, used for Hall-effect measurements, Figure S2; four terminal electrical transport properties of Sample #2, Figure S3; Hall-effect in multi-layered $ReS_2$ FETs, Figure S4; scaling analysis at the metal insulator transition for a tri-layered crystal, Figure S5; micrographs and height profile for two multi-layered $ReS_2$ FETs, Figure S6; two-terminal transport properties for a $ReS_2$ field-effect transistor, Figure S7; transport properties of Sample #5 and its two-terminal field-effect mobility Figure S8; determination of the Schottky barrier height, Figure S9; table S1 including the extracted sample dependent field-effect mobilities

**AUTHOR INFORMATION**
Corresponding Author *E-mail: balicas@magnet.fsu.edu.

Notes: The authors declare no competing financial interest.

**ACKNOWLEDGEMENTS**
This work was supported by the U.S. Army Research Office MURI Grant No. W911NF-11-1-0362. J. L. acknowledges the support by NHMFL UCGP No. 5087. Z.L .and D.S. acknowledge the support by DOE BES Division under grant no. DE-FG02- 07ER46451. The NHMFL is supported by NSF through NSF-DMR-0084173 and the State of Florida. H. T. acknowledges the support from the National Science Foundation (EFRI-1433311). A. M., R. R. N., and M. D. acknowledge the support from the US Army Research Laboratory (ARL) Director's Strategic Initiative (DSI) program on interfaces in stacked 2D atomic layered materials. V. D. is supported by NSF through Grant No. DMR-1410132. The views and conclusions contained in this document are those of the authors and should not be interpreted as representing official policies, either expressed or implied, of the Army Research Laboratory or the U.S. Government. The U.S. Government is authorized to reproduce and distribute reprints for Government purposes notwithstanding any copyright notation herein.